# Fetal Ultrasound Image Segmentation for Measuring Biometric Parameters Using Multi-Task Deep Learning

Zahra Sobhaninia, Shima Rafiei, Ali Emami, Nader Karimi, Kayvan Najarian,
Shadrokh Samavi, S.M.Reza Soroushmehr

*Abstract*— Ultrasound imaging is a standard examination during pregnancy that can be used for measuring specific biometric parameters towards prenatal diagnosis and estimating gestational age. Fetal head circumference (HC) is one of the significant factors to determine the fetus growth and health. In this paper, a multi-task deep convolutional neural network is proposed for automatic segmentation and estimation of HC ellipse by minimizing a compound cost function composed of segmentation dice score and MSE of ellipse parameters. Experimental results on fetus ultrasound dataset in different trimesters of pregnancy show that the segmentation results and the extracted HC match well with the radiologist annotations. The obtained dice scores of the fetal head segmentation and the accuracy of HC evaluations are comparable to the state-of-the-art.

## I. Introduction

Ultrasound (US) imaging is a safe non-invasive procedure for diagnosing internal body organs. Ultrasound imaging as compared to other imaging tools, such as computed tomography (CT) and magnetic resonance imaging (MRI), is cheaper, portable and more prevalent [1]. It helps to diagnose the causes of pain, swelling, and infection in internal organs, for evaluation and treatment of medical conditions [2]. Ultrasound imaging has turned into a general checkup method for prenatal diagnosis. It is used to investigate and measure fetal biometric parameters, such as the baby's abdominal circumference, head circumference, biparietal diameter, femur and humerus length, and crown-rump length. Furthermore, the fetal head circumference (HC) is measured for estimating the gestational age, size and weight, growth monitoring and detecting fetus abnormalities [3].

Despite all the benefits and typical applications of US imaging, this imaging modality suffers from various artifacts such as motion blurring, missing boundaries, acoustic shadows, speckle noise, and low signal-to-noise ratio. This makes the US images very challenging to interpret, which requires expert operators. As shown in US image samples of

Z. Sobhaninia, S. Rafiei, A. Emami, and N. Karimi are with the Department of Electrical and Computer Engineering, Isfahan University of Technology, Isfahan 84156-83111, Iran.
S. Samavi is with the Department of Electrical and Computer Engineering, Isfahan University of Technology, Isfahan 84156-83111, Iran and also with the Department of Emergency Medicine, University of Michigan, Ann Arbor, 48109 U.S.A.
K. Najarian and S.M.R. Soroushmehr are with the Department of Computational Medicine and Bioinformatics and the Michigan Center for Integrative Research in Critical Care, University of Michigan, Ann Arbor, 48109 U.S.A.

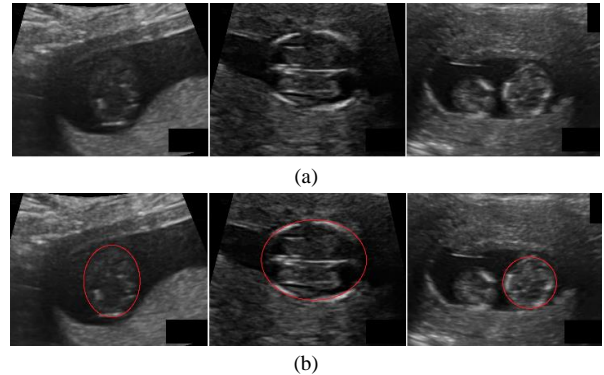

Figure 1. Samples of ultrasound fetal head dataset[1] a) Original images b) Ground truth provided by a radiologist (red borders).

Fig. 1(a) these images are noisy and blurry with incomplete shapes; furthermore, the fetal skull is not visible enough to detect in the first trimester.

In the last decade, automatic methods for fetal biometric measurements have been investigated. Development of these automated methods has improved the work flow efficiency by reducing the examination time and number of steps necessary for standard fetal measurements [3].

Past studies have used various methods for HC measurement such as randomized Hough transform [4], semi-supervised patch based graphs[5], multilevel thresholding circular shortest paths [6], boundary fragment models[7], Haar-Like features [8], active contouring [9], or compound methods such as [10] which apply Haar-like features to train a random forest classifier in order to locate the fetal skull. Then, HC was extracted by using Hough transform, dynamic programming and ellipse fitting. Although these methods provided reassuring results, they were assessed on small datasets of particular pregnancy trimesters.

Recently, deep convolutional neural networks (DCNN) have rapidly become a compelling choice for several image processing tasks such as classification, object detection, segmentation, and registration [11]. More recent researches on fetal ultrasound image analysis focus on using DCNN. For instance, Cerrolaza et al. [12] applied fully convolutional networks for skull segmentation in fetal 3D US images. They proposed a two-stage convolutional neural network which incorporated additional contextual and structural information into the segmentation process. Another research applied a cascade U-Net network to estimate the biometric parameters of the fetal abdominal area [13].

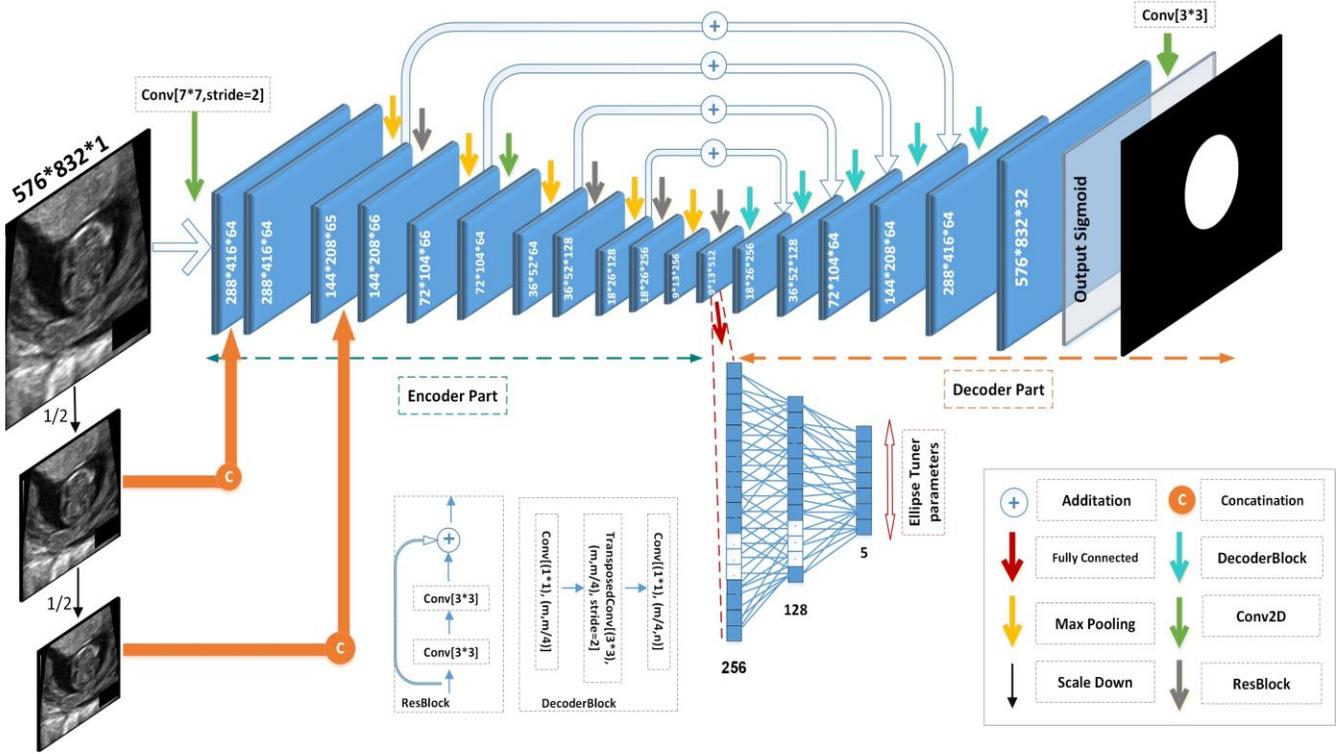

Figure 2. Overview of the proposed network architecture based on Link-Net

In this paper, we propose a multi-task deep network based on the structure of Link-Net [14] which was originally developed for semantic segmentation. We utilize the Link-Net capability for segmentation of the fetal US images. Our experimental results demonstrate that multi-task learning leads to remarkable improvement over a single-task network and produces more accurate segmentation results. The rest of this paper is structured as follows: Section 2 elaborates on the proposed method and explains the compound loss function. Section 3 discusses the experimental results. We conclude the paper in Section 4 by a brief discussion about the proposed system.

## II. PROPOSED METHOD

Our proposed network is an end to end Multi-Task network based on Link-Net architecture (MTLN) with multi-scale inputs. Fig. 2 illustrates a block diagram of the proposed system, including two main modules: a segmentation network and an Ellipse Tuner. Two different loss functions are defined for the two modules, which improve the training process for the whole network and leads to overall better performance. During the training phase, the network parameters are trained by back-propagation of the combinatorial loss gradients through different entry points:

$$L_T = \alpha_1 L_{Seg} + \alpha_2 L_{ET} \qquad (1)$$

where $L_{Seg}$ and $L_{ET}$ represent loss functions of segmentation and Ellipse Tuner, respectively. $\alpha_1$ and $\alpha_2$ define weights of the loss functions. Details of the system modules are explained in sections A and B.

### A. Segmentation Network

The proposed network is a modified version of Link-Net with multi-scale inputs, which is applied for fetal head segmentation. As shown in Fig. 2 MTLN takes 2D ultrasound image in three scales, which are fed into different layers. The first half of the network contains encoding blocks called ResBlock [15], which comprise of convolutional and pooling layers with a residual link. To provide a multi-scale structure, we have concatenated first and second feature maps of the network with the down-sampled versions of the input image. The second half of the network is composed of decoder blocks that are responsible for up-sampling the feature maps and building up the final segmentation output. As shown in Fig. 2, some skip-connections connect corresponding encoders and decoders of similar dimensions. These skip-connections help preserve feature-map details which might be lost throughout the encoding and decoding process. Use of skip-connections leads to more accurate segmentation of the boundaries.

The segmentation loss function, $L_{Seg}$, is defined by (2) based on the sum of cross-entropy (E) and Dice metric, magnified on the boundaries of the fetal head, using weight map $w$:

$$L_{Seg} = w * (E + Dice) \qquad (2)$$

$$E = -\sum_{x,y \in Z^2}(\log p_c(x,y)) \qquad (3)$$

$$Dice = 1 - \frac{2 * (G \cap S)}{|G| + |S|} \qquad (4)$$

where G is ground truth, S is the network segmentation mask and $(x, y)$ are the pixel coordinates. The weighting map $w$ in (2) is introduced in [16] for improving the training process. It magnifies the loss on boundaries of fetal head during training by emphasizing on edges, as defined below:

$$w(x) = 1 + \omega_0 \cdot \exp \frac{d(x)}{2\sigma^2} \quad (5)$$

where $d(x)$ represents the distance between the pixel $x$ and ground truth boundaries, $\sigma$ is the variance of the Gaussian kernel and $\omega_0$ is a predefined constant.

## B. Ellipse Tuner

Since all the ground truth shapes, provided by the radiologist, have elliptical shapes (c.f. Fig. 1(b)), we investigate ellipse parameters for assessment of the fetal head. For this purpose, we exploit the rich feature maps in the middle of the segmentation network, between the encoder and decoder sections, for estimating the ellipse parameters.

As shown in Fig. 2, the extracted features are fed into three Fully Connected (FC) layers for tuning ellipse parameters. The five outputs of the FC network are expected to estimate ellipse parameters which represent fetus head location, shape, etc. Furthermore, the proposed FC layers implicitly contribute to segmentation performance and improve the accuracy of the segmentation results, by refining the feature layers to symbolize an ellipse shape.

The loss function, used for training the Ellipse Tuning network, is the mean squared error (MSE) of predicted ellipse parameters as compared to the ground truth:

$$L_{ET} = \frac{1}{n} \sum_{i}^{n} (P_{Pred}(i) - P_{gt}(i))^2 \quad (6)$$

where $P_{Pred}$ is the output of Ellipse Tuner and $P_{gt}$ is parameters vector from ground truth. The ellipse parameters vector used for optimization includes the center coordinates $(c_x, c_y)$, diameters and the angle between the small diameter and the $y$ axis. An ellipse can be defined as the locus of all points that satisfy the equation:

$$\frac{(x - c_x)^2}{a^2} + \frac{(y - c_y)^2}{b^2} = 1 \quad (7)$$

where $(c_x, c_y)$ represent the ellipse center coordinates, $a$ and $b$ are the radiuses along the $x$ and $y$ axes. Parametric representation of ellipse in (8) may be derived by drawing two tangential circle, as shown in Fig. 3:

$$(x, y) = (a \cos(t), b \sin(t)) \quad 0 \le t < 2\pi \quad (8)$$

where parameter $t$ is the angle of the line passing through points A and B with x-the axis.

## III. EXPERIMENTAL RESULTS

The proposed method is implemented in python and Tensorflow and trained over 200 epochs using stochastic gradient descent with momentum (learning rate = 0.001). Training time on NVIDIA GeForce GTX 1080 Ti was about 15 hours. Table I presents the values of hyper-parameters used for training.

TABLE I. HYPER-PARAMETERS VALUE

| Parameter | $\alpha_1$=1 | $\alpha_2$=2 | $w_0$=30 | $\sigma$=10 |
|---|---|---|---|---|

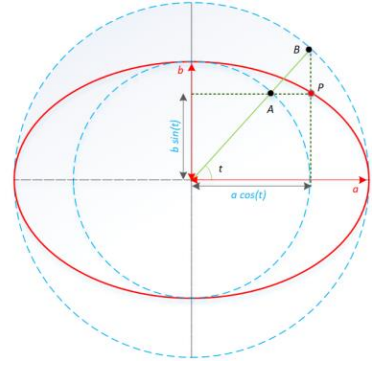

Figure 3. The construction of points based on the parametric equation and the interpretation of parameter t [17].

## A. Dataset

We collected 999 two-dimensional ultrasound images of HC from the database of Department of Obstetrics of the Radboud University Medical Center, Nijmegen, the Netherlands. For this study, we only used fetus samples without any growth abnormalities [10]. The size of ultrasound images is (800,540), and scale of pixel sizes is in the range 0.052 mm to 0.326 mm.

## B. Augmentation

For data augmentation, we transform each image by horizontal and vertical flipping, as well as using fixed rotations from -60 to 60 in steps of 20 degrees. In some cases, a rotation transform destroys the fetal head area by moving the head outside the rotated image. Hence, we remove the corrupted images from the training set and keep images with a complete fetal head. We generated 8823 augmented US images in total. The dataset was randomly split into (75%, 25%) for training and test. We keep 10% of the training set as validation data.

## C. Evaluation

We utilize $DSC$ (Dice Similarity Coefficient), DF (Difference), ADF (Absolute Difference) and HD (Hausdorff Distance) [10] for evaluation of MTLN system performance:

$$DSC = \frac{2 * (Area_S \cap Area_R)}{|Area_S| + |Area_R|} \quad (9)$$

$$DF = HC_P - HC_{GT} \quad (10)$$

$$ADF = |HC_P - HC_{GT}| \quad (11)$$

where $Area_S$ is the ground truth area and $Area_R$ is the area extracted from the segmentation network. $HC_P$ represents the calculated perimeter from segmentation result and $HC_{GT}$ is the ground truth fetal head circumference. The Hausdorff Distance (HD) is defined as:

$$H(S, R) = \max(h(S, R), h(R, S)) \quad (12)$$

$$h(S, R) = \max_{s \in S} \min_{r \in R} ||s - r|| \quad (13)$$

where $S = \{s_1, \ldots, s_q\}$ demonstrates pixels from the output segmentation results and $R = \{r_1, \ldots, r_q\}$ shows the pixels from ground truth.

To assess the effect of multi-task learning and Ellipse Tuner in the proposed network, we trained a single task network (without the FC layers) and compared against MTLN. We observed that the DSC score of single task network was 92.67 ± 2.70 and it improves to 96.84 ± 2.89. Furthermore, as shown in Fig. 4 the FC layers of MTLN leads to smoother and cleaner elliptic segmentation results.

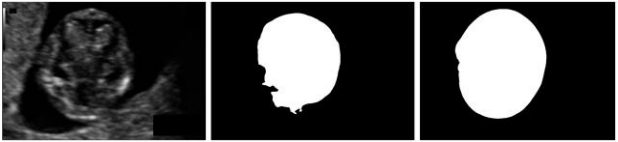

Figure 4. Comparison of segmentation results. From left to right:
1) Original image, 2) Result without FC, 3) MTLN Segmentation result.

Table II demonstrates MTLN results against Heuvel *et al.* [10] on the test dataset. The proposed system outperforms the competitor in terms of HD and ADF, while Our DSC score is also comparable to theirs.

TABLE II. QUANTITATIVE COMPARISON BASED ON SCORING CRITERIA

| Method | Dsc Score % | DF(mm) | ADF(mm) | HD (mm) |
|---|---|---|---|---|
| Heuvel et al. [10] | **97.0±2.8** | **0.6±4.3** | 2.8±3.3 | 2.0±1.6 |
| MTLN (Ours) | 96.84 ± 2.89 | 1.13±2.69 | **2.12 ± 1.87** | **1.72 ±1.39** |

For visual purposes, Fig. 5 demonstrates the segmentation results of the proposed system against Heuvel et al. [10]. The second column presents the segmentation results of [10], and the third column is the MTLN results. The red borders in both columns show the ground truth. The green and blue regions are the segmentation results of the two systems. As shown in Fig. 5, the segmentation results of our system are comparable to [10], but in some cases provides more reliable results.

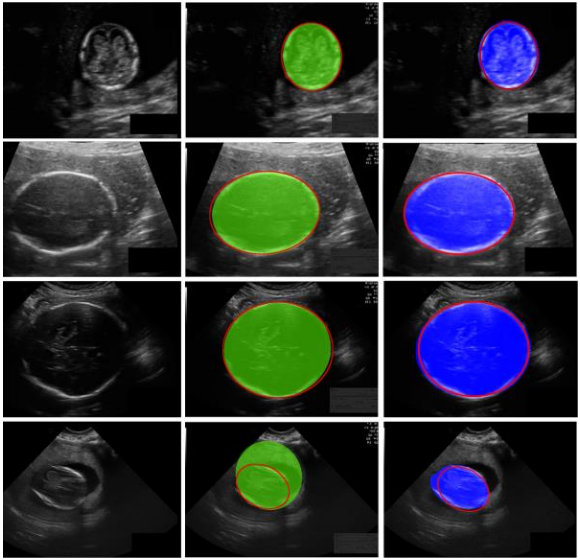

Figure 5. Segmentation results of fetal head (ground truth is red) From left to right: 1) Original image, 2) Green: results of [10], 3) Blue: MTLN results.

## IV. CONCLUSION

In this paper, we presented a Multi-Task deep network based on Link-Net structure with multi-scale inputs, for segmentation and estimation of fetal head circumference in 2D ultrasound images. Our proposed network was trained on 999 images and was evaluated on an independent test set which included data from all trimesters. The proposed system incorporated an Ellipse Tuner based on fully connected networks. We demonstrated that the performance of a multi-task network is better than the single-task network with no Ellipse Tuner and leads to smoother and cleaner elliptical segmentation results.